# Bridge Damage Detection using a Single-Stage Detector and Field Inspection Images

Chaobo Zhang*, C.C. Chang & Maziar Jamshidi
Hong Kong University of Science and Technology, Clear Way Bay, Hong Kong

## Abstract

*Detecting concrete surface damages is a vital task for maintaining the structural health and reliability of highway bridges. Currently, most of these tasks are conducted manually which could be cumbersome and time-consuming. Recent rapid advancement in convolution neural network has enabled the development of deep learning-based visual inspection techniques for detecting multiple structural damages. However, most deep learning-based techniques are built on two-stage, proposal-driven detectors and using less complex image data, which is not promising to promote practical applications and integration within intelligent autonomous inspection systems. In this study, a faster, simpler single-stage detector is proposed based on YOLOv3 for detecting multiple concrete damages of highway bridges. To realize this, a bridge damage dataset consisting of field inspection images is prepared. These images have large variations in damage appearance and monitoring scene and are labeled with four types of concrete damages: crack, pop-out, spalling and exposed rebar. The original YOLOv3 is further improved by introducing a novel transfer learning method, Batch Renormalization and Focal Loss. The improved YOLOv3 is evaluated in terms of average precision and speed. The results show that the improved YOLOv3 has a detection accuracy of up to 80%. Its performance is about 13% better than the original YOLOv3.*

## 1. Introduction

Highway bridges are an integral part of the transportation systems that form the backbone of the modern metropolis. Their safety and serviceability are vital to society and always need to be closely monitored and maintained. For this reason, highway bridges are inspected by certified professionals periodically and after any major events. On one hand, the maintenance and rehabilitation of aging bridges have become a common problem in many places including the US [1, 2]. On the other hand, for a rapidly-developing region like Hong Kong with over 1,300 concrete bridges [3], although the bridge inventory is newer than those in Europe and North America, they carry a very high traffic volume [4]. It is necessary to detect defects appearing on these bridges in early stages to prevent any further losses in their structural capacity and durability.

Surface defects are the most observable indicator of possible structural deterioration or damage. Among all non-destructive evaluation (NDE) techniques, visual inspection is most practiced. Besides, low-cost and high-quality digital cameras contributed to the recent growing incorporation of image-based techniques into the structural health monitoring (SHM) framework. It is noteworthy, however, that the internal condition of a structural element cannot be assessed by merely relying on visual techniques. Other in-depth methods must be brought into the process for a comprehensive inspection. Nevertheless, surface defects are a good measure for the general condition of a structural member and are still a key part of many condition assessment manuals [5].

Manual inspection of bridges can be costly, labor intensive, dangerous in locations with low accessibility, and subjective. These drawbacks initiated the emergence of computer vision-based inspection and robotic inspection platforms, e.g. unmanned aerial vehicle [6, 7]. Once visual data is collected from a bridge, there is an entire suite of image processing techniques (IPTs) to extract relevant information for defect detection, classification, and assessment [5]. The types of damages considered were mostly focused on concrete cracking and were extended to minor extent spalling/delamination and rusting. For crack detection, basic edge detection methods like the Haar wavelet method were shown to be reliable [8]. The performance of thresholding algorithms for noisy image data however was questionable [9]. In more advanced methods, the extracted features of images are classified using machine learning techniques in a more automated way and with less sensitivity to image noise. These machine learning-based approaches have been utilized for detecting various types of defects including cracks [10-12], spalling [13, 14], and corrosion [15]. The efficiency and robustness of traditional IPTs were significantly improved by applying machine learning techniques. However, these methods are still based on handcrafted low-level features and require pre- and post-processing. These drawbacks make them less applicable on images with large variations, such as those acquired from robotic platforms.

Lately, IPTs based on deep learning have gained a lot of attention. Convolutional neural networks (CNNs) enhance the capability of computer vision techniques in object detection and classification. Unlike a traditional neural network, a CNN learns the appropriate features automatically from the training data. This is a significant advantage because CNNs do not depend on hand-engineered features, and no prior image processing or feature extraction is necessary. In terms of network architecture, the CNN-based object detectors can be divided into two main categories: two-stage and single-stage detectors. In general, two-stage detectors are more accurate, but single-stage detectors are faster. Due to the accuracy issue, most studies in the field of deep learning-based structural inspection are based on two-stage detectors. Recent improvements on single-stage detectors, however, have allowed them to achieve comparable or even better accuracy than two-stage detectors [16, 17].

---

*czhangbd@connect.ust.hk

The motivation of the study is to investigate the performance of the state-of-the-art single-stage object detector (YOLOv3) for identifying multiple classes of defects in bridge inspection images, and to evaluate the trade-offs between accuracy and speed. The speed of the detection process can become important as the number of images increases. This is particularly true for inspection videos where processing every frame can become overwhelmingly slow. On the other hand, in a boarder scope, high-speed detection is essential for the development of real-time robotic inspection systems which seems to be the future direction of the industry [18].

The aim of this study is twofold: first, to train and configure the original YOLOv3 for defect detection. Second, to improve the original architecture and evaluate its reliability and performance. The main contributions of the article are as follows:

*- A dataset of 2,206 concrete bridge damage images was prepared and annotated. These images are actual inspection images acquired from the Hong Kong Highways Department. Therefore, the variations in damage scale and appearance in the images are such that they represent actual conditions.*

*- To improve the original architecture, a new transfer learning method is proposed to tackle the problem with training the network with a small dataset. Besides, Batch Renormalization and Focal Loss are employed as refinements to enhance the accuracy of YOLOv3.*

*- Comprehensive validation experiments are conducted to evaluate the accuracy and speed performance of the original and improved YOLOv3.*

The rest of the article is organized as follows. Section 2 discusses the related works and Section 3 provides an overview of the YOLOv3 network architecture. In Section 4, the improvements for original architecture are described. Section 5 describes the procedures for generating database and date augmentation. Experimental results are presented in Section 6, and concluding remarks are given in Section 7.

## 2. Related work

**Convolutional neural network:** Owing to its excellent capability, researchers utilized the power of CNN for structural damage detection. In the simplest form, CNN can be used to classify each pixel as either damaged or intact. Such approach was employed for crack segmentation by classifying each pixel [19, 20]. To find the extent of a defect in a large image, a simple approach is to raster scan the image with a fix-sized sliding window and flag the windows that the classifier detected. Following this approach, Cha et al. [21] proposed a deep CNN model to classify whether each image patch contains a concrete crack. In another study, Yang et al. [22] fine-tuned the VGG-16 model [23] to classify two types of defect: cracks and spalling. Applying CNN to damage classification was shown to achieve a high level of accuracy. However, the challenge of such sliding window-based approach is to find the proper window size when dealing with defects of various scales. Moreover, this approach comes with a high computational cost, since CNN classifier must be applied many times for every single window in each image.

**Two-stage detectors:** To improve the efficiency of detecting and locating multiple objects, Region-based CNN (R-CNN) uses Selective Search [24] to generate region proposals instead of sliding windows to find objects in an image [25]. R-CNN is a two-stage architecture that incorporates region proposal as an intermediate step. However, the region proposals generation step using the selective search is still slow and has limited accuracy because generating object proposals through such an external method is not effective for some small and thin objects. Also, due to the significant duplicate computation from overlapping regions, the R-CNN approach can become rather inefficient and time-consuming. Later, to overcome this critical issue, Ren et al. [26] propose the Faster R-CNN, which uses a region proposal network (RPN) to replace the traditional handcrafted proposals generation step. The Faster R-CNN offers speed and accuracy improvements over its predecessors by sharing computation and using neural networks to propose regions. These two-stage detectors have recently been used to detect structural defects. For example, Kim et al. [27] combine R-CNN with morphological post-processing to detect and quantify cracks in a concrete bridge. However, the detected crack region from R-CNN is fragmented and accuracy is affected by the extracted potential crack regions. In another study, Cha et al. [28] use the Faster R-CNN architecture to detect five types of structural surface damage in concrete and steel. They show that they can train their model to achieve high accuracy for a dataset of images with distinctive features. Li et al. [29] also employed a modified Faster R-CNN network to identify three types of concrete defects and their dataset included various background information and relatively small defects. All these studies are based on two-stage detectors incorporating region proposal as the intermediate step, thus they still fall short of real-time performance. Also, the pipeline of training these two-stage approaches is complex and requires a lot of time.

**Single-stage detectors:** Due to the above drawbacks, single-stage detectors were proposed as an end-to-end network that unifies object classification and localization in a single convolution network. There are two popular state-of-the-art models in this category: SSD (Single Shot MultiBox Detector) [30] and YOLO (You Only Look Once) [31]. Both SSD and YOLO remove the proposal-generating step and simultaneously predict multiple bounding boxes and class probabilities; therefore, their processing speed is considerably fast. The speed of the detection process can become important as the number of images increases. While they are definitely faster than two-stage detectors, some researchers found that directly locating objects using such single-shot style networks arguably compromises the accuracy of the detection [29, 32]. In particular, the earliest version of YOLO struggled with small objects and could hardly be generalized to objects with unusual aspect ratios or configurations. This indicates a significant number of localization errors. To remedy these shortcomings, Redmon and Farhadi [17, 33] enhance the performance of the network by introducing several improvements, including applying Batch Normalization [34] and using good priors of anchor boxes to predict bounding boxes in YOLOv2; and employment of Feature Pyramid Network [35] and Darknet-53 backbone in YOLOv3. They suggest that the latest version (YOLOv3) could achieve a comparable accuracy as the two-stage detectors (i.e., Faster R-CNN), without scarifying its computational efficiency. To the best of our knowledge, only very few studies have investigated the efficiency and performance of single-stage detectors for structural inspection. In a recent publication, Maeda et al. [36] applied SSD to detect road surface damages in real-time. They demonstrated that SSD could obtain relatively high accuracy for some classes of defect. In this paper, a YOLOv3-based single-stage detector is proposed and validated for detecting multiple defects in a concrete bridge.

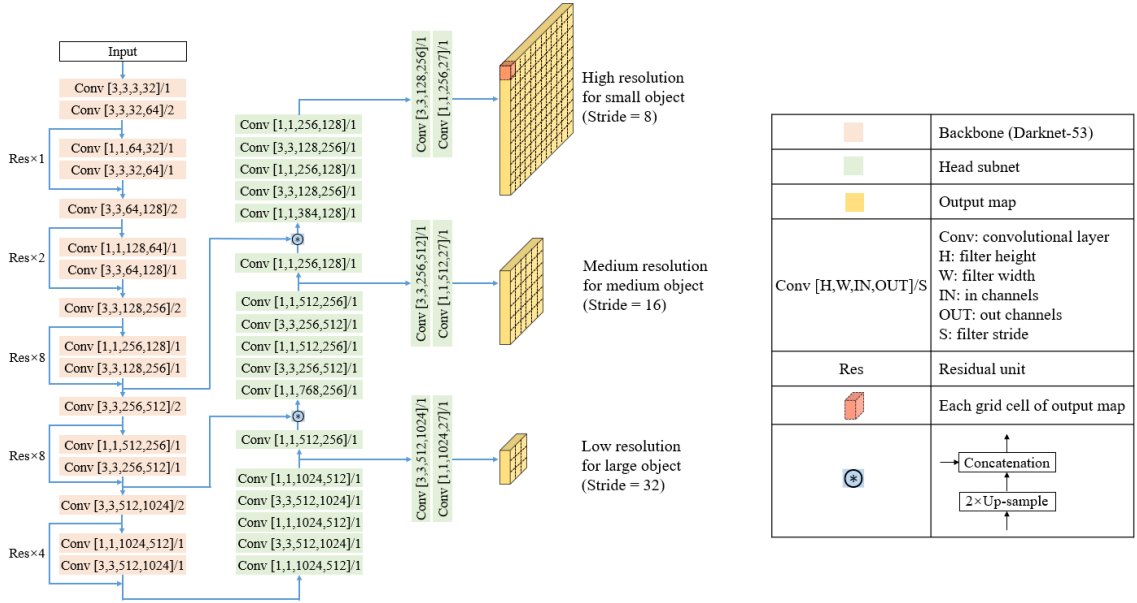

Figure 1. Architecture of YOLOv3 with multi-scale predictions.

## 3. Overview of YOLOv3 architecture

YOLOv3 is a single-stage CNN consisting of one backbone and one head subnet (see Figure 1). The backbone is responsible for computing the convolutional feature maps over an entire input image and its performance is usually evaluated by classification accuracy using the ImageNet dataset [37]. The head subnet is built on top of the backbone to perform classification and bounding box regression and then output the predicted results. The details of the YOLOv3 framework is explained in this section.

**Backbone:** The backbone of the YOLOv3 framework is called Darknet-53 which contains 23 residual units [38]. Each residual unit contains one $3 \times 3$ and one $1 \times 1$ convolutional layer. At the end of each residual unit, an element-wise addition is performed between the input and output vectors. Batch Normalization (BN) [34] is used after each convolutional layer followed by the Leaky ReLU [39] activation function. Also, the down-sampling step is performed in five separate convolutional layers with a stride of 2. The backbone Darknet-53 is pre-trained on ImageNet dataset and performs on par with state-of-the-art classifiers but with less computational cost (classification accuracy is 0.1% higher and speed is 1.5 times faster than the ResNet-101 according to [17]).

**Head subnet:** The head subnet of YOLOv3 adopts a Feature Pyramid Network (FPN) [35] to detect objects at three different scales. In brief, FPN augments a standard convolutional network with a top-down pathway and lateral connections such that the network efficiently constructs a rich, multi-scale feature pyramid from a single resolution input image. Each level of the pyramid can be used for detecting objects at a different scale. The lower resolution feature maps have larger strides that leads to a very coarse representation of the input image, which is assigned for large object detection. While the higher resolution feature maps have more fine-grained features and is used for small object detection. YOLOv3 builds FPN on top of the backbone architecture and constructs a pyramid with down sampling strides 32, 16, and 8. It uses concatenation to perform the merging step in lateral connections instead of element-wise addition used in the original FPN paper by Lin et al. [35].

**Anchors:** Unlike Faster R-CNN, YOLOv3 implements k-means clustering on training set ground-truth boxes to automatically find good priors of anchors. The good priors are box dimensions (width and height) of anchors that lead to larger overlap between anchors and ground-truth objects. The default number of clusters in YOLOv3 is set to be 9. By sorting the area of 9 clusters and then dividing them evenly across scales, specific feature maps learn to be responsive to particular scales of the objects. Hence, an anchor box with a smaller scale is assigned to a feature map with a higher resolution. During training, a ground-truth box needs to be assigned to a particular anchor box and train the network accordingly. YOLOv3 only assigns one anchor box for each ground-truth object. Thus, only the anchor box that has the highest IoU with the ground-truth box will be responsible for predicting an object. This leads to specialization among the bounding box predictors [31].

**Prediction and loss function:** Instead of predicting location offsets as in most object detectors [26, 30, 40], YOLOv3 predicts location coordinates relative to the location of the grid cell. This bounds the ground truth to fall between 0 and 1, so the logistic activation is used to constrain the network's predictions to fall within this range. In total, the network predicts 1 confidence score ($t^o$), 4 class probabilities ($p$), and 4 coordinates ($t^x, t^y, t^w, t^h$) for each of three bounding boxes at each grid cell on three different-resolution output feature maps. If the cell is offset from the top left corner of the image by ($c^x, c^y$) and the bounding box prior has width $p^w$, and height $p^h$, then the predictions correspond to

$$\begin{aligned} b^x = \sigma(t^x) + c^x, \quad b^y = \sigma(t^y) + c^y, \\ b^w = p^w e^{t^w}, \quad b^h = p^h e^{t^h}, \\ C = \sigma(t^o), \quad p^{class} = softmax(p) \end{aligned} \quad (1)$$

where $b^x$ and $b^y$ are the center coordinates, and $b^w$ and $b^h$ are the width and height of the predicted bounding box, respectively; $C$ is the object confidence; and $p^{class}$ is the probability of being each damage type. Unlike the independent logistic classifiers in YOLOv3, here since the damage classes are mutually exclusive in the bridge damage dataset, the softmax classifier [41] is used to transform the input values in a multi-class probability distribution. With these definitions, YOLOv3 minimizes an objective function which is a weighted sum of the confidence loss, classification loss, and localization loss. The cross entropy loss [42] is used to measure the performance of confidence and class predictions whose output is a probability value between 0 and 1. For the localization loss, sum-squared loss is used as the predicted values need to regress to a specific target value not necessarily between 0 and 1. Note that only the anchor box which has the highest IoU with the ground-truth box will be responsible for predicting an object. Hence, the loss function only penalizes classification errors and localization errors if that anchor box is responsible for the ground-truth box.

## 4. Improved YOLOv3

Training a deep network requires large annotated image datasets to achieve high predictive accuracy. However, in many domains like structural damages, acquisition of such data is difficult and labeling them is costly and labor intensive. Also, limited computer memory may result in a small training batch size, which diminishes the effectiveness of batch normalization. In this section, some improvements are made for original YOLOv3 to improve its efficiency for civil and practical applications.

### 4.1. Transfer learning

Instead of training the CNN from scratch, it is common to pre-train a CNN model on a very large dataset and then use the pre-trained weight either as an initialization or a fixed feature extractor for the task of interest. Yosinski et al. [43] showed that transferring features and then fine-tuning them offer better results compared with the case of freezing transferred feature layers; moreover a better performance could be achieved with more transferred feature layers. The transfer learning (TL) technology is commonly used to initialize the backbone of an object detector, where the backbone weights are usually pre-trained on the standard ImageNet classification dataset which is publicly accessible [16, 17], [26, 30]. TL technology has been widely used and has played an extremely important role in deep learning-based structural damage recognition [29, 44-46]. To apply TL for YOLOv3, the backbone of YOLOv3 is initialized with pre-trained Darknet-53 weight on ImageNet dataset and all new convolution layers without pre-trained weight are initialized with Xavier initialization [47]. This is the common practice of weight initialization for an object detector. Note that we investigated the TL using the pre-trained weight as a fixed feature extractor (freezing transferred feature layers during training) and found that it did not work well for this study. This has two plausible reasons: (1) there is a large difference between the ImageNet classification dataset and damage detection dataset; and (2) the later part of the pre-trained Darknet-53 weight is more specific and not very general for common detection task. This effect may be more severe when performing the merging step in the lateral connection of FPN.

Inspired by the fact that better performance could be achieved with more transferred feature layers, a novel transfer learning method was developed. Instead of using a damage dataset to directly fine-tune all layers, a larger common object dataset extracted from COCO dataset [48] is used to fine-tune the network firstly. Then, restoring this fully pre-trained weights for all layers and fine-tuning all layers again using damage dataset. In this method, the object instance for a fully pre-trained weight usually falls within a specific scale range. Differences in the object scale between pre-trained weight and damage datasets may result in a large domain-shift when fine-tuning from a pre-trained network. This makes the optimization process difficult. This phenomenon was also indicated in YOLOv2 [33] and SNIP [49]. In order to reduce such domain-shift between pre-trained weight and damage datasets while finetuning the network, the COCO classes with the closest distribution of object scales and aspect ratios to the damage dataset should be extracted from COCO dataset to perform the first fine-tuning step. Also, feature maps from different levels within a network are known to have different effective receptive field size [50], which only takes up a fraction of the full theoretical receptive field, and this effective receptive field evolves during training. As we assign a specific scale of anchors to each level (i.e. smaller scale for higher resolution feature map), specific feature maps learn to be responsive to particular scales of the objects. It is sensible that the object scales in pre-trained weights and custom datasets are better to be similar in terms of each level of feature maps. Hence, the default anchors for damage dataset was used when fine-tuning on the extracted COCO classes. This can help to make the object scale for each output feature maps to be similar between fully pre-trained weight and damage dataset.

### 4.2. Batch Renormalization

It is well-known that normalizing the input data makes training faster. The change in the distributions of layers' inputs presents a problem because the layers need to continuously adapt to the new distribution. Batch Normalization (BN) [34] is adopted in YOLOv3 to normalize the previous output by subtracting the batch mean and dividing by the batch standard deviation. However, simply normalizing the layer data may change what the layer can represent. To address this, BN introduces a pair of parameters, $\gamma$ and $\beta$, to scale and shift the normalized value. These two parameters are learned along with the original model parameters and they restore the representation power of the network. The formula for a BN layer is

$$y_i = \gamma \left( \frac{x_i - \mu_B}{\sqrt{\sigma_B^2 + \epsilon}} \right) + \beta \quad (2)$$

where, $x_i$ and $y_i$ represent the input feature and the output value in each hidden layer before activations; $\mu_B$ and $\sigma_B^2$ are the mini-batch mean and variance computed along each dimension (channel) of input $x_i$; and $\epsilon$ is a small real number added to variance to avoid dividing by zero.

However, BN will break when the mini-batch mean ($\mu_B$) and standard deviation ($\sigma_B$) diverge frequently from the mean and standard deviation over the entire training set. Also, at inference time, the moving averages of mean and standard deviation (as an estimate of the statistics of the entire training set) is used to do the normalization step. Naturally, if the mean and standard deviation during training and testing are different, the testing results may be worse. This can happen when the mini-batch samples are biased, or more commonly, when batch size is small due to limited computer memory. For instance, in this study, the computer can only afford a batch size of 2. Batch Renormalization [51] (BR) tackles the issue of differing statistics at train and inference time head-on. In BR, the normalization step at inference time and training time are related by an affine transformation as follows

$$y_i = \gamma \left( \frac{x_i - \mu_B}{\sqrt{\sigma_B^2 + \epsilon}} \times r + d \right) + \beta,$$
$$r = \frac{\sigma_B}{\sigma} \in \left[ \frac{1}{r_{max}}, r_{max} \right], \quad (3)$$
$$d = \frac{\mu_B - \mu}{\sigma} \in [-d_{max}, d_{max}]$$

where $r$ and $d$ are treated as constants; $\mu$ and $\sigma$ are the current moving mean and standard deviation, respectively; and $r_{max}$ and $d_{max}$ are two hyperparameters introduced to control the transition between BN and BR. Using mini-batch statistics ($\mu_B$ and $\sigma_B$) and affine transformation at training time and moving averages at inference time ensures that the output of BR is the same during both phases.

### 4.3. Focal Loss

During training YOLOv3, there is an imbalance between foreground and background confidences (e.g. 1:8000 for a $512 \times 512$ image with 2 objects). Easily classified negatives comprise the majority of the confidence loss and dominate the gradient. Also, the class imbalance between different damages in the training data is obvious as the number of exposed rebar and crack (72%) is much larger than the other two damages (28%). The confidence and class imbalance results in a large domain-shift during training. This imbalance causes two problems: (1) training is not very efficient as most locations are easy negatives that provide no useful learning signal; and (2) the damage class with a larger dataset may overwhelm training and lead to degenerate models. To solve these imbalance problems, Focal Loss (FL) [16] is applied to the original loss function. FL addresses the imbalance by using modulating factor to down-weight inliers (easy examples) in such a way that their contribution to the total loss is small even if their number is large. In other words, the FL performs training on a sparse set of hard examples. For confidence loss, the FL applied for a Cross Entropy with sigmoid outputs is as follows.

$$FL(y, \hat{y}) = \begin{cases} -(1-y)^\gamma \log(y) & , \text{ if } \hat{y} = 1 \\ -y^\gamma \log(1-y) & , \text{ otherwise} \end{cases} \quad (4)$$

And for classification loss, the FL applied for a Cross Entropy with softmax outputs is as follows

$$FL(y, \hat{y}) = -(1-y)^\gamma \log(y) \quad , \text{ if } \hat{y} = 1 \quad (5)$$

where $FL$ represents the Focal Loss function; $y$ and $\hat{y}$ are the respective predicted and ground-truth values; $(1-y)^\gamma$ and $y^\gamma$ are the modulating factors; and $\gamma$ is the focusing parameter that smoothly adjusts the rate at which easy examples are down-weighted. When $\gamma = 0$, $FL$ is equivalent to Cross Entropy, and as $\gamma$ increases the effect of the modulating factor is likewise increased (in this study $\gamma = 2$ in all experiments).

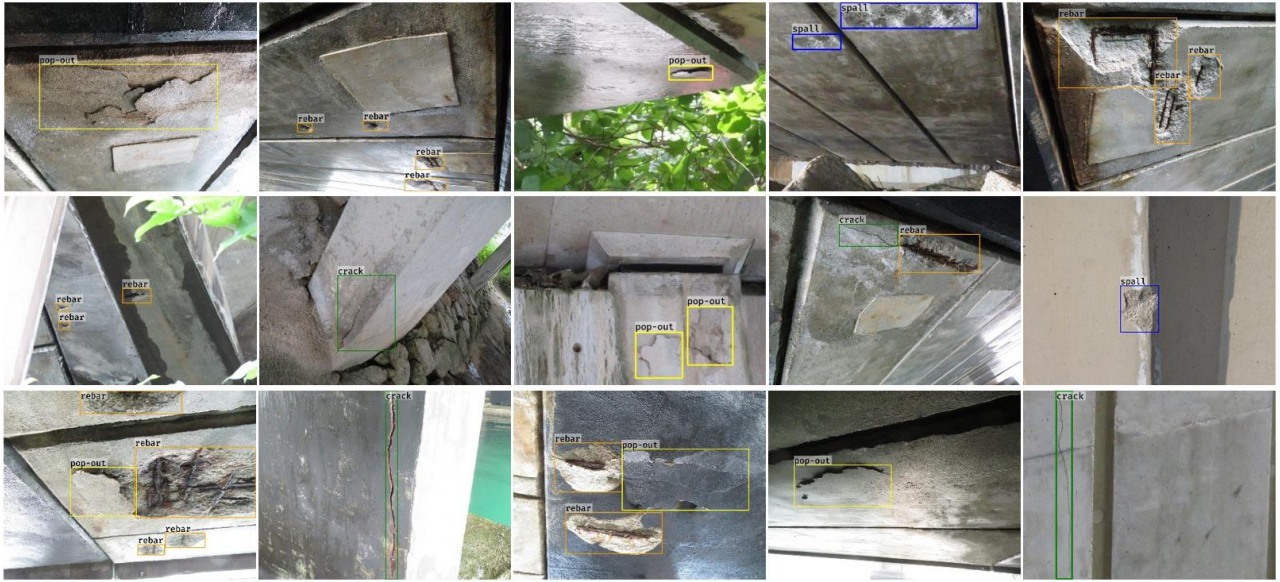

Figure 2. Sample images with ground-truth boxes and class labels.

## 5. Dataset

### 5.1. Dataset and anchors

An image dataset containing a total of 2,206 inspection images of highway concrete bridges in Hong Kong was established for this study. Among them, 75% of the images with 1,280 × 960 pixels resolution were acquired from the Hong Kong Highways Department and the rest with 4,000 × 3,000 pixels resolution were taken by the authors. The graphical image annotation tool LabelImg [52] was used to label damage objects including their damage types and bounding box coordinates. Four different types of damage are considered in this study: crack, pop-out, spalling and exposed rebar. Spalling refers to the type of damage where the concrete surface has peeled off but rebars are not exposed, while pop-out is used to indicate those damages where the concrete cover is still in place. Figure 2 shows some examples of the annotated images with different types of damages.

Large variation across object scales, especially the challenge of detecting very small objects stands out as one of the factors behind the difference in performance of object detectors. The median object area relative to the image in PASCAL VOC [53], and COCO dataset are around 8.5% and 1.2%, respectively. Comparison of detection results between COCO and PASCAL VOC dataset indicates that most objects are small, and large variation in scale makes the damage detection task more difficult [26, 30, 33]. For the bridge damage dataset, the median object area relative to the image is about 2.5%, hence most damage objects shown in image is larger than the common objects in COCO dataset but smaller than that in PASCAL VOC dataset. Moreover, there is a class imbalance between different damages. The percent of the number of damage objects per category is about 35% for crack, 15% for pop-out, 13% for spalling and 37% for exposed rebar.

Multi-scale training with seven image sizes: MS = {416, 448, 480, 512, 540, 572, 608} is performed to give the network of the ability to predict well across a variety of input dimensions. To obtain the prior anchors, k-means clustering is also considered on these seven different input sizes. By sorting the area of 9 anchors and separating them evenly across 3 prediction scales, the higher resolution feature map learns to be responsive to the objects with smaller scales. For the bridge damage dataset, the nine clusters were: (29×22), (30×95), (97×37), (39×267), (105×101), (290× 59), (227×139), (126 × 282), (411 × 209). These nine anchors can offer an average IoU of 0.583 with ground-truth objects and 70.4% of the ground-truth boxes have an anchor with IoU larger than 0.5.

### 5.2. Data augmentation

Data augmentation is a way to increase the performance of object detector and can reduce the probability of overfitting on small dataset. Considering the actual operation of bridge inspection and the potential application of robotic systems, the following four aspects were considered for data augmentation. Firstly, there is image blurring or degradation because of camera movement during the image capturing process. The motion blur is more severe for camera installed on a flight system (e.g. UAV) due to wind effect. The shift

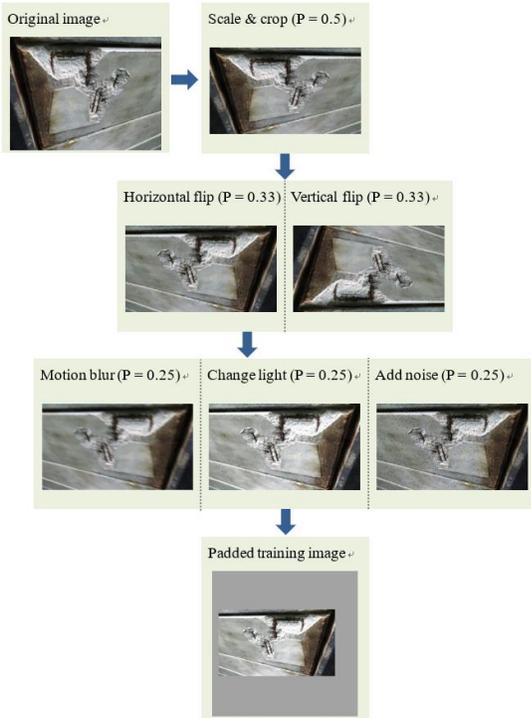

Figure 3. Data augmentation procedure.

in the image can be simulated by a linear motion blur function (e.g. [54, 55]), which includes motion direction and length. Secondly, due to random illuminations of the scenes, the real inspection images have large brightness variations. The brightness of an image can be changed by changing the pixel values of 'Lightness' in HLS (hue, saturation, lightness) color space. Thirdly, digital images are often prone to the contamination of impulse noise due to the deficiency in the hardware of communication systems, electro-magnetic inference, etc. Salt-and-pepper noise can be used to model defects in the charge-coupled device (CCD) image sensor or in the transmission of the image, for which a certain amount of the pixels in the image are either black or white. Last but not least, due to varying viewpoint and camera-damage distance in real-world application, the damage instances have a very large variation in the scale and aspect ratio. For this effect, randomly scaling and cropping the image can make the model more robust to various input object sizes and shapes. These four aspects help to generate training examples that cover the span of image variations and help the network to be less sensitive to changes in these properties. The improved performance of the proposed data augmentation procedure was discussed later. In summary, each training image is augmented by the following sequence (see Figure 3),

1. Randomly scale and crop with a probability of 1/2. Note that in randomly cropping, only keep the training image if all ground-truth boxes are shown in this cropped area.

2. Horizontally or vertically flip the image with a probability of 1/3.

3. Randomly manipulate image by applying the motion blur, changing brightness or adding salt-and-pepper noise with a probability of 1/4.

4. Input a padded training image to the network.

## 6. Experiments and discussions

The interpreted high-level programming language Python and the open-source software library TensorFlow are used to perform experimental study. Training and testing were performed on a PC running the Ubuntu 17.04 operating system with an Intel® CoreTM i7-7700 CPU and a GeForce GTX 1060 6GB GPU. The YOLOv3 network is trained using the Adam optimizer [56] with a batch size of 2 and a weight decay of 0.0001. The total number of training epochs was fixed to be 80 with a relatively large learning rate $10^{-3}$ for the first 25 epochs, $10^{-4}$ for the second 25 epochs, and $10^{-5}$ and $10^{-6}$ for the last two 15 epochs. The same data augmentation procedure and multi-scale training with seven image size MS are employed in all training YOLOv3 models.

To evaluate the network performance, the dataset was randomly split into a ratio of 8:2 with the former part as training data and the latter part as testing data. The testing data is a holdout dataset used to provide an unbiased evaluation of the final model and it was not used in the training process [57]. Average precision ($AP$) is used to evaluate the detection performance, which summarizes the precision/recall curve by calculating the area under the curve [53]. In this study, mean average precision at both IoU = 0.5 ($mAP_{50}$) and IoU = 0.75 ($mAP_{50}$) are considered for evaluating the detection accuracy. The strict IoU metric of 0.75 is used to count for some special tasks that requires a very accurate localization [48].

### 6.1. Cross-validation results

To study the performance of different transfer learning methods, normalization techniques, and loss functions, the training data of bridge damage dataset is randomly partitioned into 5 equal sized subsamples according to fivefold cross-validation principle. Each subsample has the same proportion of each damage class. Note that there is no need to evaluate all the combinations of above improvements. The transfer learning is always recommended when training network with a small dataset as it has a large effect on training and testing accuracy. And the small batch size problem for BN due to limited computer memory is generally believed to reduce some accuracy. The focal loss is considered as the final improvement that may affect or improve the accuracy. Hence, these three improvements were evaluated sequentially. For simplicity, the validation accuracy in this section is shown and evaluated with an image size of 512.

To compare the commonly used and developed TL methods, three experiments were conducted as outlined: (1) No TL: all layers are trained from scratch; (2) TL-A: the backbone is initialized with the pre-trained Darknet-53 weight and then all layers are fine-tuned using damage dataset (common practice); (3) TL-B: the backbone is initialized with the

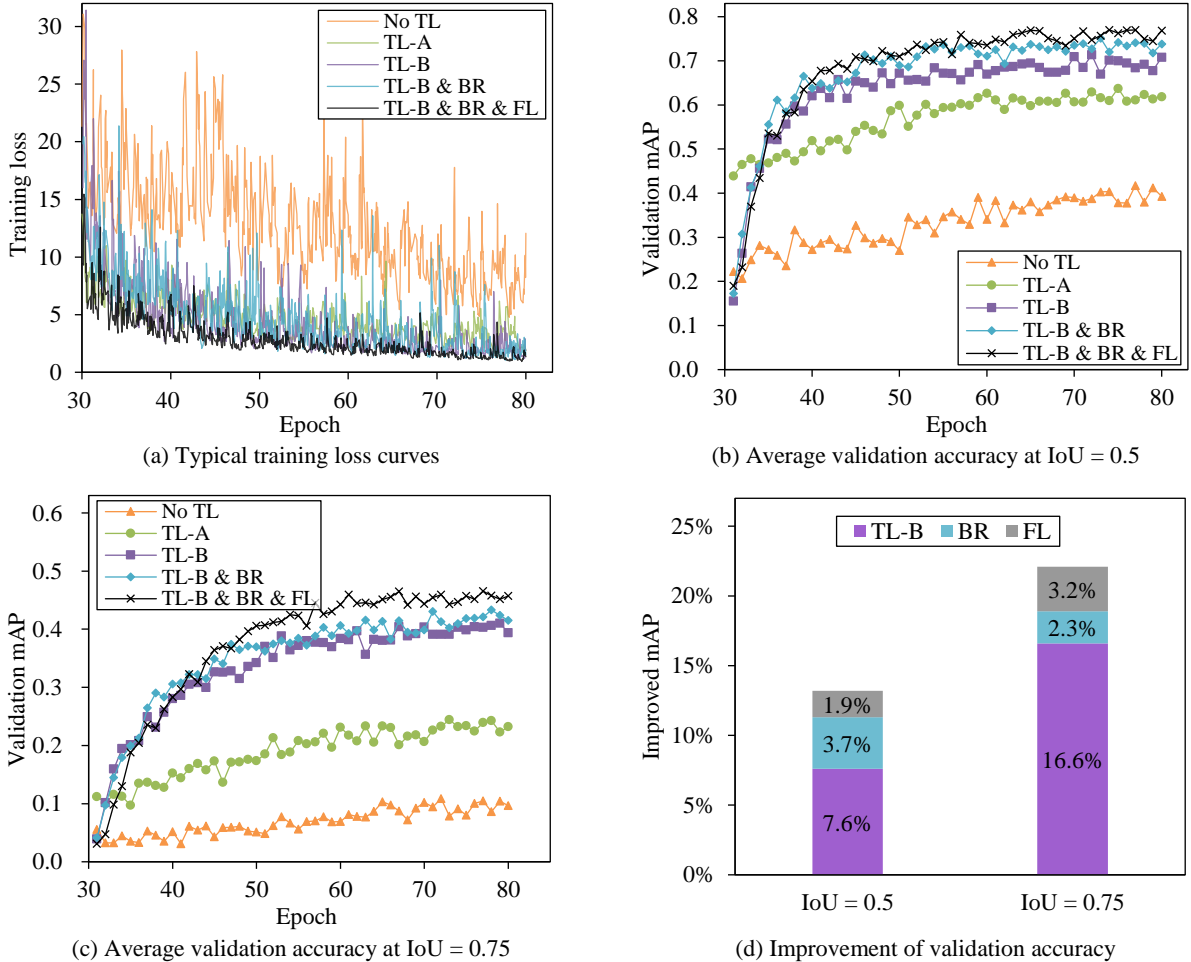

(a) Typical training loss curves

(b) Average validation accuracy at IoU = 0.5

(c) Average validation accuracy at IoU = 0.75

(d) Improvement of validation accuracy

Figure 4. Comparison between different transfer learning technologies, normalization and loss functions.

pre-trained Darknet-53 weight. Firstly, all layers are fine-tuned on the extracted four COCO classes for the first 30 epochs. Then, this fully pre-trained weight for all layers are restored and all layers are fined-tuned again using damage dataset for the remaining 50 epochs. To apply TL-B, four COCO classes with the closest distribution of object scales and aspect ratios to the damage dataset were extracted from COCO dataset to perform the first fine-tuning step. The total data number of the extracted four COCO classes (bench, fork, skateboard and truck) was 17,952, which is about nine times larger than the bridge damage dataset. Moreover, to apply BR to YOLOv3, the hyper-parameters $r_{max} = 1.5$ and $d_{max} = 0.5$ were carefully chosen to offer the best results.

Figure 4 (a) shows the comparison of typical training loss curves (with a moving average of 0.6). The validation accuracy at IoU = 0.5 and IoU = 0.75 over last 50 training epochs are shown in Figures 4(b) and (c), respectively. The validation accuracy is evaluated after each training epoch and is presented as the average value of fivefold cross-validation. Both the training loss and validating accuracy show a noisy curve and such fluctuation is more serve for the model without using TL. This instability is due to the inherent feature of Adam optimizer and the small training batch size. Also, the multi-scale training will cause some instability as the validating accuracy is evaluated on a fixed image size of 512. Comparing the three TL experiments, the models with TL converge much faster and more stably in finding the global minimum of the loss function. Results showed that the TL-B using a fully pre-trained weight from a relatively larger and similar-scale dataset offers the highest accuracy around $mAP_{50} = 71.4\%$, which is 7.6% higher than the commonly used TL-A (see Figure 4(d)). For the strict evaluation metric, the TL-B gives a $mAP_{75} = 41.0\%$, which is 16.6% higher than the TL-A. This indicates that a more accurate location prediction can be achieved using the TL-B. Also, it is obvious that network initialized with transferred weights can improve generalization performance and has much better performance than that using random weights. Comparing the validation accuracy for YOLOv3 using BN and BR, the validation accuracy $mAP_{50}$ and $mAP_{75}$ were increased by 3.7% and 2.3%, respectively. Also, the FL further gives a 1.9% to 3.2% gain on the accuracy $mAP_{50}$ and $mAP_{75}$, respectively. A possible reason for a larger increase in $mAP_{75}$ is that FL down-weights the confidence and classification loss from easy examples and thus penalizes more on the localization loss.

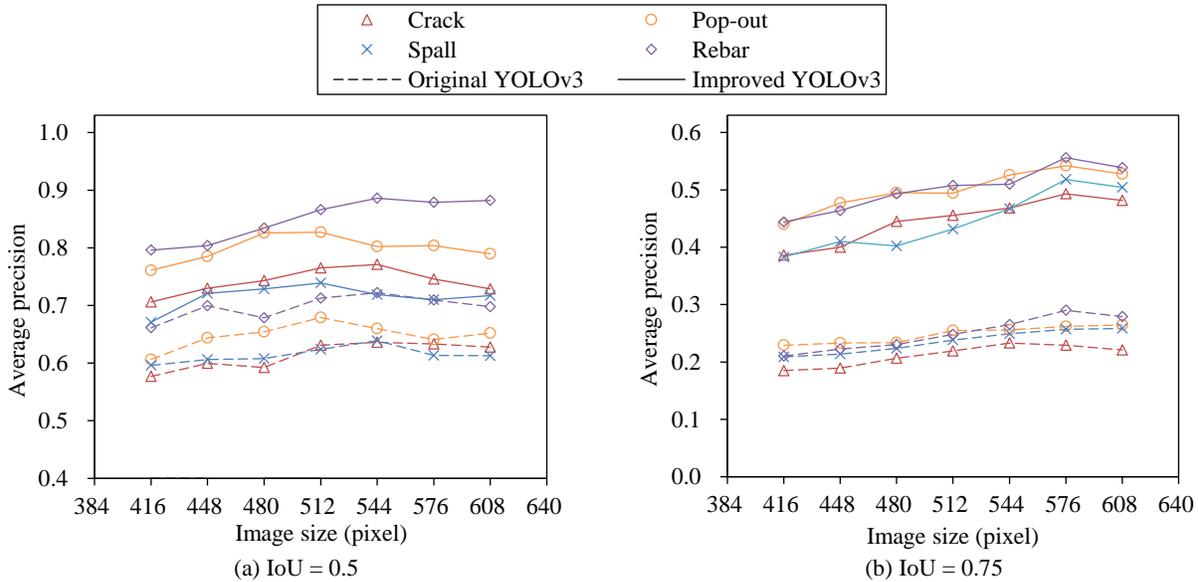

Figure 5. Testing accuracy for each class at (a) IoU = 0.5 and (b) IoU = 0.75.

In summary, a novel transfer learning method (TL-B) and some improvements (BR and FL) are proposed for YOLOv3 to increase its detection performance. We denote the YOLOv3 with TL-B and BR and FL improvements as "improved YOLOv3" and the YOLOv3 with only TL-A as "original YOLOv3". Two case studies without using the proposed data augmentation procedure were also conducted on original and improved YOLOv3. Results show that the validation accuracy $mAP_{50}$ and $mAP_{75}$ for original YOLOv3 decreases by about 18.1% and 9.6%, respectively. For improved YOLOv3, the effect of data augmentation is relatively small, which decrease the validation accuracy $mAP_{50}$ and $mAP_{75}$ by about 9.4% and 6.1%, respectively. These indicate that data augmentation is crucial for increasing the performance of object detector, especially for such small damage dataset.

### 6.2. Testing results

In this section, the testing data is used to evaluate the performance and potential application of the improved YOLOv3. Unlike the training and validation data for cross-validation that are used to fit the model and tune the model's hyperparameters, the testing data is an independent dataset employed only to assess the performance (i.e. generalization) of a fully specified detector. To obtain the testing results of YOLOv3, the model is trained again using all training data and its accuracy is evaluated on the testing data. The testing results are evaluated on different input size as that used in multi-scale training. Note that the testing performance for original YOLOv3 is also evaluated to provide reference values. Figures 5 and 6 summarize the testing accuracy for each class and the mean average precision at both IoU = 0.5 and IoU = 0.75 under different testing image sizes. For the improved YOLOv3, the $AP_{50}$ for crack, pop-out, spalling and exposed rebar at an input image size of 512 is 76.5%, 82.7%, 73.9%, and 86.6%, respectively. The $mAP_{50}$ and $mAP_{75}$ of the improved YOLOv3 is 79.9% and 47.2% respectively, which is about 13.7% and 23.2% higher than the original model at input image size of 512. The much higher increase in $mAP_{75}$ indicates that the improved YOLOv3 with the novel TL method and improvements gives an accurate location prediction under such small dataset. Comparing the average precision between different classes (see Figure 5), the exposed rebar has a higher accuracy than the other three classes. This is because the texture representing the exposed rebar tends to be more consistent compared to other damages and it has a relatively large dataset. The average precision of crack and spalling was lower than other two classes and this is mainly due to the insufficient features of crack and small scales, as well as, small dataset of spalling.

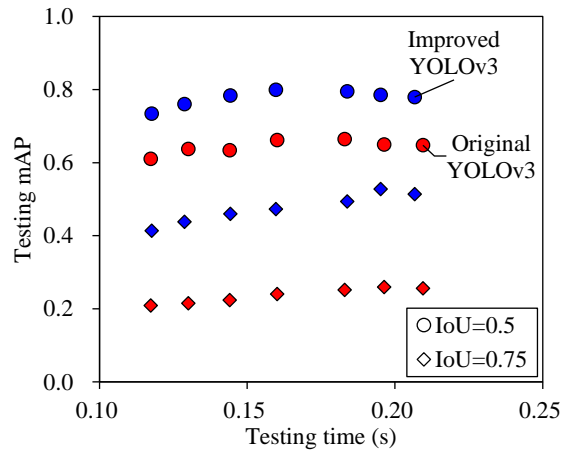

Figure 6. Detection results on testing dataset.

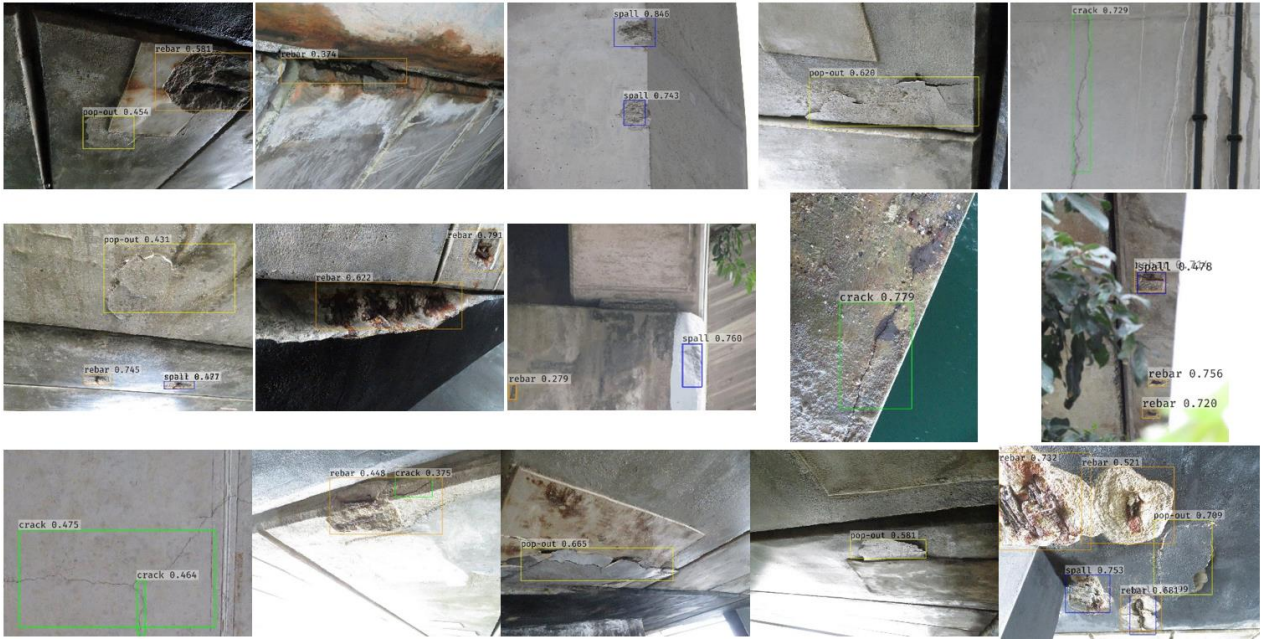

Figure 7. Detection samples from improved YOLOv3 model.

According to Figure 6, the mean average precision generally increases as the image size increases and the accuracy difference between different image sizes is not very large. This indicates that the multi-scale training makes the network capable of predicting well across a variety of input dimensions. It is, however, observed that there are a few drops in accuracy when the image size increases. This phenomenon can be attributed to: (1) the effective receptive filed on lowest resolution feature map that is responsible for detecting large objects is not large enough to predict very large objects from increased image size; (2) there is a mismatch between image size and anchor size when the image size becomes very large; (3) the possible instability from the multi-scale training.

Figure 7 shows some detection examples of the improved YOLOv3 and the testing results show a consistent performance for inspection images taken under various conditions of locations, viewpoints, camera-damage distance, lighting, blur, and background. It should be noted that the speed in Figure 6 is much slower than the reported value in original YOLOv3 paper, which can achieve a real-time detection at 20 frames per second at an image size of 608. The difference is mainly due to different GPU performance.

## 7. Conclusions

A vision-based approach for detecting multiple surface damages in concrete highway bridges is proposed based on the single-stage detector YOLOv3. To apply YOLOv3 for concrete bridge inspection, a database comprising 2,206 actual bridge inspection images is collected. The field inspection images have large variations in damage appearance and monitoring scene. They are labeled for four types of concrete damages, namely concrete crack, pop-out, spalling and exposed rebar. Testing results showed that the original YOLOv3 model can achieve a detection accuracy of $mAP_{50}$ = 66.2% and $mAP_{75}$ = 24.0% for two IoU metrics of 0.5 and 0.75, respectively.

Next, three modifications are proposed to improve the original YOLOv3. A novel transfer learning method is proposed to fine-tune the network with a fully pre-trained weight from a larger common dataset having the closest distribution of object scale and aspect ratio to the damage dataset. Cross-validation results show that applying this new transfer learning method increases the validation accuracy $mAP_{50}$ and $mAP_{75}$ by about 7.6% and 16.6%, respectively. Also, to solve the problem of low performance when training on small batch size, Batch Normalization in the original version is replaced by Batch Renormalization. This modification increases $mAP_{50}$ and $mAP_{75}$ by 3.7% and 2.3%, respectively. Lastly, the Focal Loss is introduced to solve the imbalance problem in confidence and class predictions, by down-weighting easy examples and performing training on a sparse set of hard examples. According to the validation results, the Focal Loss further increased the $mAP_{50}$ and $mAP_{75}$ by 1.9% and 3.2%, respectively. Testing results showed that the improved YOLOv3 model can achieve a detection accuracy of $mAP_{50} = 79.9\%$ and $mAP_{75} = 47.2\%$. An area for future research involves the incorporation of trained YOLOv3 models with embedded inspection systems. For such systems, with limited computational and memory resources, a small and low latency backbone like MobileNet [58] should be considered.

# References

[1] AAOSHAT (2008), *Bridging the Gap: Restoring and Rebuilding the Nation's Bridges*, American Association of State Highway and Transportation Officials, Washington DC.

[2] Adeli, H. & Jiang, X. (2008), *Intelligent Infrastructure: Neural Networks, Wavelets, and Chaos Theory for Intelligent Transportation Systems and Smart Structures*, CRC Press, Taylor & Francis, Boca Raton, Florida.

[3] Hui, M. C. & Yau, D. (2011), Major bridge development in Hong Kong, China-past, present and future, *Frontiers of Architecture and Civil Engineering in China*, **5**(4), 405–14.

[4] Yang, Y., Pam, H., Kumaraswamy, M. & Ugwu, O. (2006), Life-cycle maintenance management strategies for bridges in Hong Kong, in *Proceedings of the Joint International Conference on Construction Culture, Innovation and Management*, Dubai, 26–29.

[5] Koch, C., Georgieva, K., Kasireddy, V., Akinci, B. & Fieguth, P. (2015), A review on computer vision based defect detection and condition assessment of concrete and asphalt civil infrastructure, *Advanced Engineering Informatics*, **29**(2), 196-210.

[6] Ellenberg, A., Kontsos, A., Bartoli, I. & Pradhan, A. (2014), Masonry crack detection application of an unmanned aerial vehicle, in *Proceedings of the International Conference on Computing in Civil and Building Engineering*, 1788–95.

[7] Kim, H., Sim, S. & Cho, S. (2015), Unmanned aerial vehicle (UAV)-powered concrete crack detection based on digital image processing, in *Proceedings of 6th International Conference on Advances in Experimental Structural Engineering, 11th International Workshop on Advanced Smart Materials and Smart Structures Technology*, University of Illinois, US.

[8] Abdel-Qader, I., Abudayyeh, O. & Kelly, M. E. (2003), Analysis of edge-detection techniques for crack identification in bridges, *Journal of Computing in Civil Engineering*, **17**(4), 255–63.

[9] Iyer, S. & Sinha, S. K. (2005), A robust approach for automatic detection and segmentation of cracks in underground pipeline images, *Image and Vision Computing*, **23**(10), 921–33.

[10] Nishikawa, T., Yoshida, J., Sugiyama, T. & Fujino, Y. (2012), Concrete crack detection by multiple sequential image filtering, *Computer-Aided Civil and Infrastructure Engineering*, **27**(1), 29–47.

[11] Prasanna, P., Dana, K., Gucunski, N. & Basily, B. (2012), Computer-vision based crack detection and analysis, *SPIE Smart Structures and Materials Nondestructive Evaluation and Health Monitoring*, 834542–834542.

[12] Zalama, E., Gómez-García-Bermejo, J., Medina, R. & Llamas, J. (2014), Road crack detection using visual features extracted by Gabor filters, *Computer-Aided Civil and Infrastructure Engineering*, **29**(5), 342–58.

[13] German, S., Brilakis, I. & DesRoches, R. (2012), Rapid entropy-based detection and properties measurement of concrete spalling with machine vision for post-earthquake safety assessments, *Advanced Engineering Informatics*, **26**(4), 846–58.

[14] Dawood, T., Zhu, Z. & Zayed, T. (2017), Machine vision-based model for spalling detection and quantification in subway networks, *Automation in Construction*, **81**, 149–60.

[15] O'Byrne, M., Schoefs, F., Ghosh, B. & Pakrashi, V. (2013), Texture analysis based damage detection of ageing infrastructural elements, *Computer-Aided Civil and Infrastructure Engineering*, **28**(3), 162–77.

[16] Lin, T. Y., Goyal, P., Girshick, R., He, K. & Dollár, P. (2018), Focal loss for dense object detection, in *Proceedings of the IEEE International Conference on Computer Vision*, Los Alamitos, CA, 2999–3007.

[17] Redmon, J. & Farhadi, A. (2018), Yolov3: an incremental improvement, arXiv preprint arXiv:1804.02767.

[18] Lattanzi, D. & Miller, G. (2017), Review of robotic infrastructure inspection systems, *Journal of Infrastructure Systems*, **23**(3), 04017004.

[19] Zhang, A., Wang, K. C., Li, B., Yang, E., Dai, X., Peng, Y., Fei, Y., Liu, Y., Li, J. Q. & Chen, C. (2017), Automated pixel-level pavement crack detection on 3D asphalt surfaces using a deep-learning network, *Computer-Aided Civil and Infrastructure Engineering*, **32**(10), 805–19.

[20] Yang, X., Li, H., Yu, Y., Luo, X., Huang, T. & Yang, X. (2018), Automatic pixel-level crack detection and measurement using fully convolutional network, *Computer-Aided Civil and Infrastructure Engineering*.

[21] Cha, Y. J., Choi, W. & Büyüköztürk, O. (2017), Deep learning-based crack damage detection using convolutional neural networks, *Computer-Aided Civil and Infrastructure Engineering*, **32**(5), 361–78.

[22] Yang, L., Li, B., Li, W., Liu, Z., Yang, G. & Xiao, J. (2017), Deep concrete inspection using unmanned aerial vehicle towards cssc database, in *Proceedings of the IEEE/RSJ International Conference on Intelligent Robots and Systems*, Vancouver, BC, Canada, 24–8.

[23] Simonyan, K. & Zisserman, A. (2014), Very deep convolutional networks for large-scale image recognition, arXiv preprint arXiv:1409.1556.

[24] Uijlings, J. R., Van De Sande, Koen EA, Gevers, T. & Smeulders, A. W. (2013), Selective search for object recognition, *International Journal of Computer Vision*, **104**(2), 154–71.

[25] Girshick, R., Donahue, J., Darrell, T. & Malik, J. (2014), Rich feature hierarchies for accurate object detection and semantic segmentation, in *Proceedings of the IEEE Conference on Computer Vision and Pattern Recognition*, Columbus, OH, 580–87.

[26] Ren, S., He, K., Girshick, R. & Sun, J. (2015), Faster R-CNN: towards real-time object detection with region proposal networks, in *Proceedings of the Advances in Neural Information Processing Systems*, Montreal, 91–9.

[27] Kim, I., Jeon, H., Baek, S., Hong, W. & Jung, H. (2018), Application of crack identification techniques for an aging concrete bridge inspection using an unmanned aerial vehicle, *Sensors*, **18**(6), 1881.

[28] Cha, Y. J., Choi, W., Suh, G., Mahmoudkhani, S. & Büyüköztürk, O. (2018), Autonomous structural visual inspection using region-based deep learning for detecting multiple

[28] *damage types*, Computer-Aided Civil and Infrastructure Engineering, **33**(9), 731–47.
[29] Li, R., Yuan, Y., Zhang, W. & Yuan, Y. (2018), Unified vision-based methodology for simultaneous concrete defect detection and geolocalization, *Computer-Aided Civil and Infrastructure Engineering*, **33**(7), 527–44.
[30] Liu, W., Anguelov, D., Erhan, D., Szegedy, C., Reed, S., Fu, C. Y. & Berg, A. C. (2016), SSD: single shot multibox detector, in *Proceedings of European Conference on Computer Vision*, Springer, 21–37.
[31] Redmon, J., Divvala, S., Girshick, R. & Farhadi, A. (2016), You only look once: unified, real-time object detection, in *Proceedings of the IEEE Conference on Computer Vision and Pattern Recognition*, Las Vegas, Nevada, 779–88.
[32] Wang, L., Lu, Y., Wang, H., Zheng, Y., Ye, H. & Xue, X. (2017), Evolving boxes for fast vehicle detection, arXiv preprint arXiv:1702.00254.
[33] Redmon, J. & Farhadi, A. (2017), YOLO9000: better, faster, stronger, in *Proceedings of the IEEE Conference on Computer Vision and Pattern Recognition*, Honolulu, HI.
[34] Ioffe, S. & Szegedy, C. (2015), Batch normalization: accelerating deep network training by reducing internal covariate shift, arXiv preprint arXiv:1502.03167.
[35] Lin, T. Y., Dollár, P., Girshick, R., He, K., Hariharan, B. & Belongie, S. (2017), Feature pyramid networks for object detection, in *Proceedings of the IEEE Conference on Computer Vision and Pattern Recognition*, Honolulu, HI, 936–44.
[36] Maeda, H., Sekimoto, Y., Seto, T., Kashiyama, T. & Omata, H. (2018), Road damage detection and classification using deep neural networks with smartphone images, *Computer-Aided Civil and Infrastructure Engineering*.
[37] Deng, J., Dong, W., Socher, R., Li, L., Li, K. & Fei-Fei, L. (2009), Imagenet: a large-scale hierarchical image database, in *Proceedings of the IEEE Conference on Computer Vision and Pattern Recognition*, Miami, FL, 248–55.
[38] He, K., Zhang, X., Ren, S. & Sun, J. (2016), Deep residual learning for image recognition, in *Proceedings of the IEEE Conference on Computer Vision and Pattern Recognition*, Las Vegas, NV, 770–78.
[39] Maas, A. L., Hannun, A. Y. & Ng, A. Y. (2013), Rectifier nonlinearities improve neural network acoustic models, in *Proceedings of the 30th International Conference on Machine Learning*, Atlanta, Georgia, 1–6.
[40] Girshick, R. (2015), Fast R-CNN, in *Proceedings of the IEEE International Conference on Computer Vision*, Santiago, Chile, 1440–48.
[41] Bishop, C. M. (2006), *Pattern Recognition and Machine Learning (Information Science and Statistics)*, Springer-Verlag, New York.
[42] Goodfellow, I., Bengio, Y. & Courville, A. (2016), *Deep Learning*, MIT Press. Available at: http://www.deeplearningbook.org/, accessed March 2018.
[43] Yosinski, J., Clune, J., Bengio, Y. & Lipson, H. (2014), How transferable are features in deep neural networks, in *Proceedings of the Advances in Neural Information Processing Systems*, Montreal, 3320–28.
[44] Gopalakrishnan, K., Khaitan, S. K., Choudhary, A. & Agrawal, A. (2017), Deep convolutional neural networks with transfer learning for computer vision-based data-driven pavement distress detection, *Construction and Building Materials*, **157**, 322–30.
[45] Kucuksubasi, F. & Sorguc, A. (2018), Transfer learning-based crack detection by autonomous UAVs, arXiv preprint arXiv:1807.11785.
[46] Gao, Y. & Mosalam, K. M. (2018), Deep transfer learning for image-based structural damage recognition, *Computer-Aided Civil and Infrastructure Engineering*, **33**(9), 748–68.
[47] Glorot, X. & Bengio, Y. (2010), Understanding the difficulty of training deep feedforward neural networks, *Journal of Machine Learning Research*, **v9**, 249–56.
[48] Lin, T. Y., Maire, M., Belongie, S., Hays, J., Perona, P., Ramanan, D., Dollár, P. & Zitnick, C. L. (2014), Microsoft COCO: common objects in context, in *Proceedings of the European Conference on Computer Vision*, Springer, Berlin, 740–55.
[49] Singh, B. & Davis, L. S. (2018), An analysis of scale invariance in object detection–SNIP, in *Proceedings of the IEEE Conference on Computer Vision and Pattern Recognition*, Salt Lake City, Utah, 3578–87.
[50] Luo, W., Li, Y., Urtasun, R. & Zemel, R. (2016), Understanding the effective receptive field in deep convolutional neural networks, in *Proceedings of Advances in Neural Information Processing Systems*, Barcelona, Spain, 4898–906.
[51] Ioffe, S. (2017), Batch renormalization: towards reducing minibatch dependence in batch normalized models; 2017. arXiv preprint arXiv:1702.03275.
[52] Tzutalin (2015), LabelImg. Git code. https://github.com/tzutalin/labelImg.
[53] Everingham, M., Van Gool, L., Williams, C. K., Winn, J. & Zisserman, A. (2010), The pascal visual object classes (VOC) challenge, *International Journal of Computer Vision*, **88**(2), 303–38.
[54] Moghaddam, M. E. & Jamzad, M. (2006), Linear motion blur parameter estimation in noisy images using fuzzy sets and power spectrum, *EURASIP Journal on Advances in Signal Processing*, 2007(1), 068985.
[55] Hallermann, N. & Morgenthal, G. (2014), Visual inspection strategies for large bridges using Unmanned Aerial Vehicles (UAV), in *Proceedings of 7th IABMAS, International Conference on Bridge Maintenance, Safety and Management*, Shanghai, 661–67.
[56] Kingma, D. P. & Ba, J. (2014), Adam: a method for stochastic optimization, arXiv preprint arXiv:1412.6980.
[57] Brownlee, J. (2017), *What is the Difference Between Test and Validation Datasets*, Machine Learning Process. Available at: https://machinelearningmastery.com/difference-testvalidtion-datasets/, accessed July 2018.
[58] Sandler, M., Howard, A., Zhu, M., Zhmoginov, A. & Chen, L. (2018), MobileNetV2: inverted residuals and linear bottlenecks, in *Proceedings of the IEEE Conference on Computer Vision and Pattern Recognition*, Salt Lake City, Utah, 4510–20.